\documentclass[osajnl, twocolumn, showpacs, superscriptaddress, 10pt]{revtex4-1}
\usepackage{graphicx}
\usepackage{amsmath}
\usepackage{color}
\usepackage{sidecap}

\graphicspath{{plot/}{Figure/}}

\def\re{\mathrm{Re}}
\def\im{\mathrm{Im}}

\begin{document}

\title{A continuously tunable modulation scheme for precision control of optical cavities with variable detuning}

\author{W. Yam}
\address{LIGO Laboratory, Massachusetts Institute of Technology, Cambridge, MA 02139.}
\author{E. Davis}
\address{Department of Physics, University of California, Berkeley, CA 09125.}
\author{S. Ackley}
\address{University of California, San Francisco, CA 94110.}
\author{M. Evans}
\address{LIGO Laboratory, Massachusetts Institute of Technology, Cambridge, MA 02139.}
\author{N. Mavalvala}
\address{LIGO Laboratory, Massachusetts Institute of Technology, Cambridge, MA 02139.}

\begin{abstract}

We present a scheme for locking optical cavities with arbitrary detuning many line widths from resonance using an electro-optic modulator that can provide arbitrary ratios of amplitude to phase modulation. We demonstrate our scheme on a Fabry-Perot cavity, and show that a well-behaved linear error signal can be obtained by demodulating the reflected light from a cavity that is detuned by several line widths.

\end{abstract}

\maketitle


\section{Introduction}
	
High finesse optical cavities that are operated detuned from resonance are ubiquitously present in a variety of applications that include cavity optomechanics \cite{kippenbergRMP2013}, laser interferometer gravitational wave detectors \cite{LIGO1992, LIGO2009, LIGO2015, Virgo2008, GEO2010}, cavity QED experiments \cite{waltherRPP2006,rohdeJOSAB2002}, cold atom microscopy \cite{brahmsNatPhys2011}, and ion trapping \cite{haeffnerPhysRep2008}, to name a few.

Acquiring low noise, high bandwidth signals for locking optical cavities that are detuned from resonance by an arbitrary amount is a challenge. The Pound-Drever-Hall (PDH) locking method \cite{PDH1983} is perhaps the most common technique used for locking cavities on resonance. PDH error signals are generated by phase modulating the light incident on the cavity at RF frequencies that typically place the modulation sidebands outside the line width of the cavity, and demodulating the light reflected from the cavity. This is an extremely efficient and powerful technique for cavities operating on resonance, but the error signal becomes nonlinear and even changes sign when the cavity is detuned from resonance, making it unsuitable for use at detunings exceeding a small fraction of a line width. Conversely, linear error signals for cavities that are detuned by half a line width can be obtained by amplitude modulating the incident light. An alternative method of locking a detuned cavity using the approximately linear regime of the intensity of the transmitted light -- ``side of fringe'' or ``DC'' locking -- is simple to implement, but is prone to noise couplings and offset drifts. Again the linear regime of the signal is a small fraction of a cavity line width, and the signal vanishes on resonance. Certainly, neither PDH nor side of fringe methods are useful for cavities that are detuned from resonance by several line widths. We demonstrate a technique where an admixture of amplitude modulation (AM) and phase modulation (PM) yields useful error signals for cavities that are on resonance or continuously detuned by many line widths. 

Cusack et al. proposed the ``universally tunable modulator (UTM)'' where an electro-optic amplitude modulator was modified to generate an admixture of AM and PM \cite{cusack2004}. They connected the inputs of the two crystals of the electro-optic modulator to separate electronic drives and showed that arbitrary ratios of AM and PM could be generated by controlling the relative phase difference between the two drives. Here we use a similar implementation of a dual input electrooptic modulator to generate smoothly varying error signals for a cavity operated on resonance or continuously detuned by several line widths.
	
\section{Theory}
		
To determine the error signals from an optical cavity driven with an arbitrary combination of phase and amplitude modulated light, we review the theory of modulated light from the UTM, and then derive expressions for the error signals for arbitrary detuning of the cavity.  

The total field of a light beam modulated at a single frequency, $\omega_{m}$, can be written as	
\begin{equation}
E = E_{0} \left[1 + \re[\tilde{A} \exp \left(i \omega_{m} t \right)] \right] \exp\left(i \,\re[\tilde{P} e^{i \omega_{m} t}] \right)
\label{unimod}
\end{equation}
where we have $\tilde{A} = A e^{i \phi_{A}}$ and $\tilde{P} = P e^{i \phi_{P}}$.  If we assume small modulation depths for both AM $(A \ll 1)$ and PM $(P \ll 1)$, we have	
\begin{equation}
E = E_{0} \left(1 + \re[\tilde{A} \exp \left(i \omega_{m} t \right)] + i \, \re[\tilde{P} \exp \left(i \omega_{m} t \right)] \right)
\label{unimod}
\end{equation}
Regrouping terms corresponding to the sidebands and carrier:		
\begin{equation}
E = E_{0} + E_{+} \exp \left(i \omega_{m}t \right) + E_{-} \exp \left(-i \omega_{m}t \right)
\label{csb}
\end{equation}
where we have $E_{+} = \displaystyle \frac{\tilde{A }+ i\tilde{P}}{2} E_{0}$ and $E_{-} = \displaystyle \frac{\tilde{A}^{*} + i\tilde{P}^{*}}{2} E_{0}$, respectively.

\subsection{The Universal Tunable Modulator}
		
\begin{figure}
\includegraphics[width=0.35\textwidth]{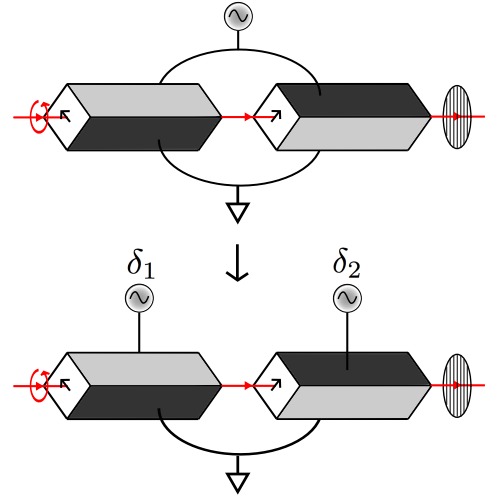}
\caption{An electro-optic amplitude modulator in normal operation (upper) and as a UTM (lower). In normal AM operation, a single input voltage source drives two crystals oriented in orthogonal directions. A UTM consists of the same two crystals wired to two independent sources. The incident light is circularly polarized.}
\label{fig:utmdiagram}
\end{figure}

The UTM of Cusack et al consists of a modified electro-optic amplitude modulator, shown in Figure \ref{fig:utmdiagram}.  An amplitude modulator consists of two identical birefringent crystals with their optic axes aligned orthogonal to each other, and aligned such that they are left and right diagonal with respect to a vertical polarizer at the output of the modulator (see Figure \ref{fig:utmdiagram}).  When the two crystals are driven exactly out of phase, the electric field along one direction is phase advanced while the orthogonal polarization is phase retarded. The average phase is zero. There is no net phase modulation from elliptical to circular polarization and back again. Filtering this oscillating polarization with a vertical polarizer produces an amplitude modulated beam.
		
When the crystals are driven by separate phasors $\tilde\delta_{1}$ and $\tilde\delta_{2}$, any modulation state within the space of AM and PM can be achieved.  The optical field exiting the UTM is given by:
\begin{equation}
\vec{E} = \tilde E_{L}  e^{i \re \left[\tilde \delta_{1}  {\rm e}^{i \omega_{m} t}\right]} \hat{L} + \tilde E_{R}  e^{i \re \left[\tilde \delta_{2}  {\rm e}^{i \omega_{m} t}\right]} \hat{R}
\end{equation}
where $\tilde E_{L} = E_{L}  {\rm e}^{i \sigma_{L}}$ and $\tilde E_{R} = E_{R}  {\rm e}^{i \sigma_{R}}$ are the electric field components along the left and right diagonal of the UTM vertical axis, and $\tilde \delta_{1} = \delta_{1}  {\rm e}^{i \sigma_{1}}$ and $\tilde \delta_{2} = \delta_{2}  {\rm e}^{i \sigma_{2}}$ are the two drive phasors. Assuming left and right diagonal field components are equal, i.e., $E_{L} = E_{R} = E_{in}/\sqrt{2}$, and the signal drives $\delta_{1,2}$ are small, we have:
\begin{equation}
\vec{E} = \frac{E_{in}}{\sqrt{2}} \left[(1 + i \re[\tilde \delta_{1}  {\rm e}^{i \omega_{m} t}]) \hat{L} +  {\rm e}^{i \sigma} (1 + i \re[\tilde \delta_{2} e^{i \omega_{m} t}]) \hat{R}\right]
\end{equation}
where $\sigma = \sigma_{R} - \sigma_{L}$. Passing through the vertical polarizer filters along the direction of $\frac{\hat{L} + \hat{R}} {\sqrt{2}}$, giving:
\begin{multline}
E_{out} =  E_{0} [1 + \frac{1}{2} \tan{\frac{\sigma}{2}} \re[(\tilde \delta_{1} - \tilde \delta_{2}) {\rm e}^{i \omega_{m} t}] +  \\
i (\frac{1}{2} \re[(\tilde \delta_{1} + \tilde \delta_{2})  {\rm e}^{i \omega_{m} t}])]
\label{modev}
\end{multline}
where $E_{0} = \displaystyle \frac{E_{in}}{2} |1 +  {\rm e}^{i \sigma}| = E_{in} \cos(\sigma / 2)$, discarding a global phase factor. By comparing equations (\ref{unimod}) and (\ref{modev}), we see that the transfer function from UTM input signals to the output modulation is:
\begin{gather}
\tilde P = \frac{1}{2} (\tilde \delta_{1} + \tilde \delta_{2})
\label{phase} \\
\tilde A = \frac{1}{2} \tan(\frac{\sigma}{2}) (\tilde \delta_{1} - \tilde \delta_{2})
\label{amplitude}
\end{gather}

Here we have assumed that left and right diagonal field components entering the UTM are equal in magnitude.  This transfer function differs from that derived by Cusack et al. by a factor of $\cos(\sigma/2)$, where the carrier suppression has been factored out. If the input field is circularly polarized, there is a symmetry between the two modulation phasors and the modulation depths are equal when driving each input separately.  
		
\subsection{Generating a cavity error signal with modulated light}

For a Fabry-Perot cavity of length $L$ with input and end mirror reflectivities $r_{1}$ and $r_{2}$, respectively, and corresponding transmittances of $t_{1}$ and $t_{2}$, the reflection coefficient, defined as the ratio of the cavity reflected field to the incident field, is
\begin{equation}
r_{cav}(\phi) = r_{1} - \frac{t_{1}^{2} r_{2}  {\rm e}^{i \phi}}{1 - r_{1} r_{2}  {\rm e}^{i \phi}}
\end{equation}
where the round-trip phase shift is given by $\phi = 2 \omega L / c$.

Rewriting this in terms of line widths of detuning from resonance, we have:
\begin{equation}
r_{cav}(n_{lw}) = r_{1} - \frac{t_{1}^{2} r_{2}  {\rm e}^{i n_{lw} \phi_{lw}}}{1 - r_{1} r_{2}  {\rm e}^{i n_{lw} \phi_{lw}}}
\end{equation}
where $n_{lw}$ is the number of line widths of detuning and $\phi_{lw} = 2 \omega_{lw} L / c$ is the phase accumulated per line width of detuning, for line width $\omega_{lw}$.

To calculate the error signal, we apply the cavity reflection coefficient to the carrier and sideband fields of the form of equation (\ref{csb}), giving a total reflected field:
\begin{multline}
E_{ref}(n_{lw}) = r_{cav}(n_{lw}) E_{0} \\
+ r_{cav}(n_{lw} + n_{m}) E_{+}  {\rm e}^{i \omega_{m} t} \\
+ r_{cav}(n_{lw} - n_{m}) E_{-}  {\rm e}^{-i \omega_{m} t}
\end{multline}
where $n_{m} = \phi_{m} / \phi_{lw}$ is the line widths of detuning accumulated by the sidebands.

The reflected power is now:
\begin{multline}
P_{ref}(n_{lw}) = |E_{ref}(n_{lw})|^2 = \\ 
\left\{\left[r_{cav}(n_{lw}) E_{0}\right]^{*}r_{cav}(n_{lw} + n_{m}) E_{+} \right. \\
+ \left. r_{cav}(n_{lw}) E_{0} \left[r_{cav}(n_{lw} - n_{m}) E_{-}\right]^{*}\right\} e^{i \omega_{m} t} \\
+ \left\{\left[r_{cav}(n_{lw}) E_{0}\right]^{*}r_{cav}(n_{lw} - n_{m}) E_{-} \right.  \\
+ \left. r_{cav}(n_{lw}) E_{0} \left[r_{cav}(n_{lw} + n_{m}) E_{+}\right]^{*}\right\} e^{-i \omega_{m} t} \\
+ (\textrm{DC Terms}) + (2 \omega_{m} \textrm{Terms})
\end{multline}
Separating out $\sin \omega_{m} t$ and $\cos \omega_{m} t$ terms and writing explicitly in terms of $\tilde A$ and $\tilde P$, we have:
\begin{multline}
P_{ref}(n_{lw}) = P_{0}[\re[\mathcal{E}(n_{lw})] \cos \omega_{m} t
     - \im[\mathcal{E}(n_{lw})] \sin \omega_{m} t] \\
+ (\textrm{DC Terms}) + (2 \omega_{m} \textrm{Terms})
\end{multline}
where $P_{0} = E_{0}^2$ and $\mathcal{E}(n_{lw})$ is defined as:
\begin{multline}
\mathcal{E}(n_{lw}) = [r_{cav}(n_{lw})]^{*} r_{cav}(n_{lw} + n_{m})(\tilde A + i \tilde P) \\
+ r_{cav}(n_{lw}) [r_{cav}(n_{lw} - n_{m})]^{*}(\tilde A - i \tilde P)
\label{errorsig}
\end{multline}
We define the $\cos \omega_{m} t$ term as being the in phase error signal, or I phase, and the $\sin \omega_{m} t$ term as being the in quadrature phase, or Q phase.	

For high finesse cavity and a modulation frequency well outside of the cavity line width and larger than the desired detuning,
\begin{equation}
 r_{cav}(n_{lw} \pm n_m) \sim 1, 
\end{equation}
such that
\begin{equation}
 \mathcal{E}(n_{lw}) \simeq 2 \Big( \re[ r_{cav}(n_{lw})] \tilde A + \im[r_{cav}(n_{lw})] \tilde P \Big)
\end{equation}
from which it is clear that the demodulated error signal from the cavity can be made to cross zero at any detuning $n_{lw}$ by setting the modulation phases such that
\begin{equation}
\frac{\im[r_{cav}(n_{lw})]}{ \re[ r_{cav}(n_{lw})]} =  \frac{-\tilde A}{ \tilde P}
\end{equation}
										
\section{Experimental Setup}

\begin{figure}[t]
\includegraphics[width=0.45\textwidth]{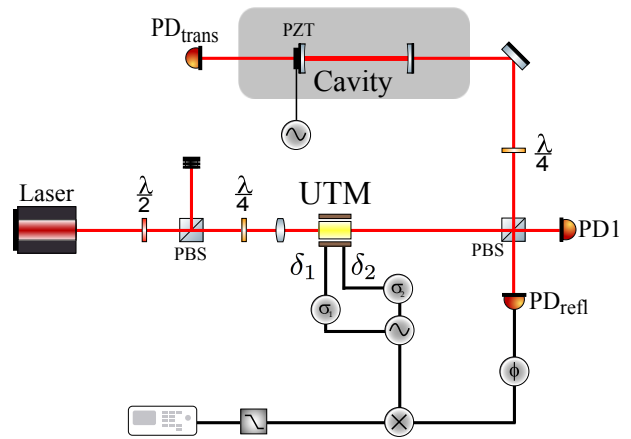}
\caption{Schematic of the experimental setup. The path of the laser beam is shown in red and electronic wiring is shown in grey.  Laser light in an optimal polarization is modulated by the UTM and incident a linear Fabry-Perot cavity. The cavity length is actuated by a PZT stack on the end mirror of the cavity. The two inputs to the UTM are driven by the two separate outputs of a two-channel signal generator. The LO port of the mixer is fed with a 25.23 MHz signal from derived from one channel of a 2-channel signal generator. The RF port of the mixer is fed with the signal coming from ${\rm PD_{ref}}$.}
\label{fig:setup}
\end{figure}
	
The experimental setup is shown in Figure \ref{fig:setup}.  Light from a 1064nm ND:YaG laser passes through a half-wave plate, a polarizing beam splitter (PBS), a quarter-wave plate, and through the UTM.  Our UTM is a modified New Focus 4140 amplitude modulator. The optical train leading to the UTM ensures that the light incident on the UTM is circularly polarized. A sample beam is incident on PD1 to directly detect amplitude modulation for diagnostic purposes.  The main beam is incident on a free space cavity that is 10 cm in length. The input and end mirrors of the cavity have amplitude reflectivity of $r_1 = 0.98 $ and $r_2 = 0.997$, respectively, which gives the cavity a finesse of 270. The cavity end mirror is attached to a PZT, enabling actuation of the cavity length.  

The cavity length was scanned through resonances by driving the PZT; the resulting cavity error signal was measured by demodulating the signal on  ${\rm PD_{ref}}$ (New Focus 1811) with a local oscillator (LO). The LO was derived from one of the two UTM drives using a power splitter.
		
The inputs to the UTM were driven separately using a signal generator that outputs two signals with an arbitrary phase between them. This ability to control the relative phases of the two signals going into the UTM is necessary to explore the entire modulation space. The relative phases of the drive signals going into the UTM were adjusted in the range from 0 to 180 degrees in 30 degree intervals.  The phase of the signal with respect to the LO was adjusted to measure both quadratures of the demodulated signal.  

\section{Results}

\begin{figure}[t]
\includegraphics[width=0.5\textwidth]{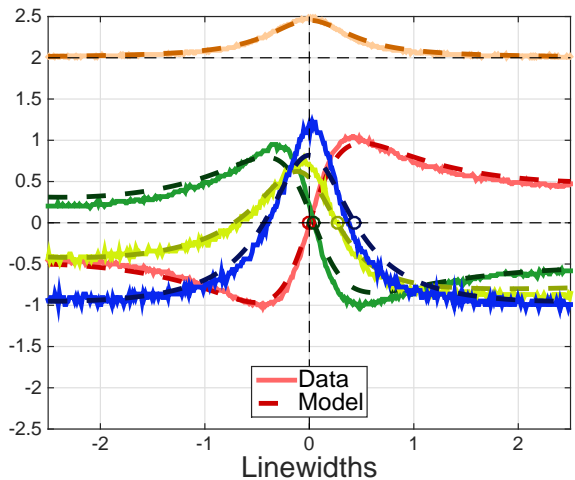}
\caption{Cavity error signals in phase (I phase) as a function of cavity length detuning from resonance. The different curves correspond to different relative phase differences of the input drives of $0^{\circ}$ (red), $60^{\circ}$ (green), $120^{\circ}$ (yellow), $180^{\circ}$ (blue). Dashed and solid lines correspond to the model and data, respectively, for each phase difference. The cavity transmission for a single fit is also overlaid at the top. The units on the y-axis are arbitrary.}
\label{fig:Iphase}
\end{figure}

Figure \ref{fig:Iphase} shows the error signals as a function of cavity detuning for different modulation states, along with the theoretical model corresponding to equation (\ref{errorsig}) above.  As we detune the cavity, we are able to generate error signals based on different modulation states ranging from pure PM to pure AM. We emphasize that only a subset of the possible modulation states are shown here.  For example, one could drive one crystal only, varying polarization of the light instead to create different modulation depths of AM and PM. This would correspond to the AM and PM modulations being in phase.  

The theoretical model was fit using five parameters: two parameters accounted for the difference in the slope and offset of the amplitude of the error signal, two other parameters accounted for the transformation from the sweep time to cavity length actuation (using a second order polynomial fit), and another parameter accounted for the zero point offset.  				
		
\begin{figure}
\includegraphics[width=0.45\textwidth]{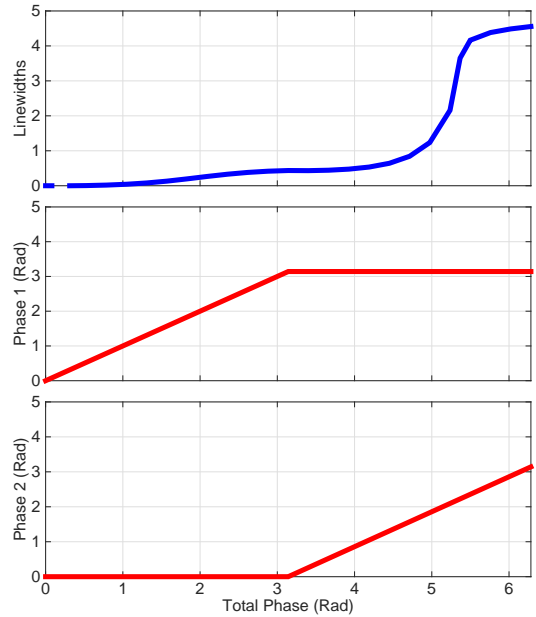}
\caption{The upper panel shows the number of line widths of detuning as a function of the phases of input drives. The phase here is defined as the sum of the phases of the input drives relative to the phase of the LO.  As shown in the middle and lower panels, the manner in which the phase is increased is important. The phase of one drive is allowed to increase to $180^{\circ}$ before increasing the other drive to the same value.}
\label{fig:lw}
\end{figure}

An important question to ask is: How far can the cavity be detuned while still yielding a useable error signal? It turns out that the if we allow the phases of the UTM drive signal vary relative to the LO phase, the cavity can be detuned past the nominal half line width of detuning using pure AM. In Figure \ref{fig:lw}, we plot the detuning (in line widths) as a function of input drive phases. We emphasize that this phase has a slightly different definition. Previously we assumed that the LO phase was locked to one of the UTM drives, hence only the phase difference was important.  Here the phase is set by first locking the phase of one of the drives to the LO, and varying the other drive from 0 to $180^{\circ}$. Next, the drive is locked at $180^{\circ}$ and the other drive is varied in the same manner. The phase in Figure \ref{fig:lw} is defined as the sum of the phases of the two drives, ranging from both drives locked to the LO to both drives $180^{\circ}$ out of phase with the LO.  We see that this method provides a way to detune the cavity by multiple line widths. Thus, the error signal generated using UTM provides a convenient way to control the amount of detuning of a cavity.
				
\section{Concluding remarks}
We have shown that we can use the vast modulation space of AM and PM provided by a universal tunable modulator to generate error signals for a cavity that can be on resonance or many line widths detuned from resonance. By varying the phases of the drive signals to the UTM we can keep the cavity locked at different detunings. Our theoretical model for the cavity error signal agrees well with our measurements of the error signals derived by fixing the drive amplitudes and varying the phase differences of the drives. This is a fraction of the modulation space available, but has proved powerful technique for locking optical cavities with arbitrary detuning from resonance. 

\begin{acknowledgments}
We thank our current and former colleagues at the MIT LIGO Lab. In particular, we are grateful to Tim Bodiya, Thomas Corbitt, Tomoki Isogai, Eric Oelker, Shannon Sankar, Denis Martynov, and Sam Waldman for their many contributions to this work. The authors gratefully acknowledge the support of the National Science Foundation and the LIGO Laboratory, operating under cooperative Agreement No. PHY-0757058
\end{acknowledgments}
		
\appendix

\section{Derivation of the UTM transfer function}
\label{utmtf}

\end{document}